\documentclass[conference]{IEEEtran}
\IEEEoverridecommandlockouts
\newcommand{\linebreakand}{%
  \end{@IEEEauthorhalign}
  \hfill\mbox{}\par
  \mbox{}\hfill\begin{@IEEEauthorhalign}
}

\usepackage{cite}
\usepackage{multirow}
\usepackage{amsmath, amssymb, amsfonts}
\usepackage{algorithmic}
\usepackage{algorithm} % Include this if you need the 'algorithm' floating environment.
\usepackage{graphicx}
\usepackage{textcomp}
\usepackage{booktabs}
\usepackage{tabularx}
\usepackage[table,xcdraw]{xcolor} % Adjust according to your document's context, e.g., Beamer.
\usepackage{url}
\urldef\publisheddoi\url{https://doi.org/10.1109/SmartNets61466.2024.10577644}
\usepackage{pgfplots}
\pgfplotsset{compat=1.17}
\pgfplotsset{
  colormap={coolwarm}{
    rgb255=(59,76,192)
    rgb255=(124,159,249)
    rgb255=(221,221,221)
    rgb255=(245,156,125)
    rgb255=(180,4,38)
  }
}

\def\BibTeX{{\rm B\kern-.05em{\sc i\kern-.025em b}\kern-.08em
    T\kern-.1667em\lower.7ex\hbox{E}\kern-.125emX}}
\begin{document}

\title{On the Performance of Malware Detection Classifiers Using Hardware Performance Counters
\thanks{\textsuperscript{\dag}These authors contributed equally to this work.}
\thanks{\copyright~2024 IEEE. Personal use of this material is permitted. Permission from IEEE must be obtained for all other uses, in any current or future media, including reprinting/republishing this material for advertising or promotional purposes, creating new collective works, for resale or redistribution to servers or lists, or reuse of any copyrighted component of this work in other works. This is the author's accepted version of the paper published in the 2024 International Conference on Smart Applications, Communications and Networking (SmartNets 2024). The final published version is available at \publisheddoi.}
}

\author{
\IEEEauthorblockN{\makebox[0.31\linewidth]{Alireza Abolhasani Zeraatkar\textsuperscript{\dag}}}
\IEEEauthorblockA{\makebox[0.31\linewidth]{\textit{ECE Department}} \\
\makebox[0.31\linewidth]{\textit{University of California}}\\
\makebox[0.31\linewidth]{Davis, United States} \\
\makebox[0.31\linewidth]{abolhasani@ucdavis.edu}}
\and
\IEEEauthorblockN{\makebox[0.31\linewidth]{Parnian Shabani Kamran\textsuperscript{\dag}}}
\IEEEauthorblockA{\makebox[0.31\linewidth]{\textit{ECE Department}} \\
\makebox[0.31\linewidth]{\textit{University of California}}\\
\makebox[0.31\linewidth]{Davis, United States} \\
\makebox[0.31\linewidth]{pkamran@ucdavis.edu}}
\and
\IEEEauthorblockN{\makebox[0.31\linewidth]{Inderpreet Kaur}}
\IEEEauthorblockA{\makebox[0.31\linewidth]{\textit{ECE Department}} \\
\makebox[0.31\linewidth]{\textit{University of California}}\\
\makebox[0.31\linewidth]{Davis, United States} \\
\makebox[0.31\linewidth]{inpkaur@ucdavis.edu}}
\and
\IEEEauthorblockN{\makebox[0.31\linewidth]{Nagabindu Ramu}}
\IEEEauthorblockA{\makebox[0.31\linewidth]{\textit{ECE Department}} \\
\makebox[0.31\linewidth]{\textit{University of California}}\\
\makebox[0.31\linewidth]{Davis, United States} \\
\makebox[0.31\linewidth]{nramu@ucdavis.edu}}
\and
\IEEEauthorblockN{\makebox[0.31\linewidth]{Tyler Sheaves}}
\IEEEauthorblockA{\makebox[0.31\linewidth]{\textit{ECE Department}} \\
\makebox[0.31\linewidth]{\textit{University of California}}\\
\makebox[0.31\linewidth]{Davis, United States} \\
\makebox[0.31\linewidth]{tsheaves@ucdavis.edu}}
\and
\IEEEauthorblockN{\makebox[0.31\linewidth]{Hussain Al-Asaad}}
\IEEEauthorblockA{\makebox[0.31\linewidth]{\textit{ECE Department}} \\
\makebox[0.31\linewidth]{\textit{University of California}}\\
\makebox[0.31\linewidth]{Davis, United States} \\
\makebox[0.31\linewidth]{hsalasaad@ucdavis.edu}}
}

\maketitle

\begin{abstract}

Malware detection using Hardware Performance Counters (HPC) has emerged as a promising solution to improve the security of computing systems as a complement to antivirus software. Hardware-based malware detectors (HMD) use Machine Learning (ML) classifiers to detect malicious application patterns. The inputs to ML classifiers are low-level performance features known as HPCs, hardware-related activity data collected from a processor at run time to profile the low-level microarchitectural behavior of an application. This paper proposes malware detection using HPCs and machine learning classifiers and highlights the effectiveness of malware detection at run-time. We use ensemble learning techniques to improve the performance of the hardware-based malware detectors, which reduces the number of necessary micro-architectural events. This improves the processor's efficiency by eliminating the need to run an application several times since a processor can measure only 2 to 8 events at a cycle. We use 18 machine-learning models along with two ensemble learning methods to evaluate the malware detection performance, creating a total of 144 different configurations. The experimental results show that the ensemble learning-based malware detection with 2 HPCs using the ensemble technique outperforms standard classifiers with 8 HPCs by up to 10\%. It also matches the performance of standard ML-based detectors that use 16 HPCs while requiring only 4 HPCs, thereby enabling effective run-time malware detection.
\end{abstract}

\begin{IEEEkeywords}
Hardware Malware Detection, Hardware Performance Counters, Ensemble Learning, Machine Learning
\end{IEEEkeywords}

\section{Introduction}
The ever-increasing demand for modern, sophisticated computing systems has resulted in the growth of critical security issues and the attempt of attackers to compromise software and hardware infrastructures. Attackers compromise systems by using emerging hardware vulnerabilities to threaten security and perform malicious activities. Attackers can harm or disrupt the system, steal the users' data, or combine all these threats. Attackers typically hide malicious code in a seemingly benign application to deceive users into downloading the malicious application. Among various forms of incentives, malware can be found everywhere, from desktop PCs to server systems, smartphones, and tablets. Malware can impair a computing system in different ways. Malware, short for malicious software, is a program or application designed to compromise the computing system without the user's consent. Malware can steal sensitive information like credit card information, login credentials, and personal information, or they can destroy files, obtain unlimited access to resources, or even seize control of the system. Hence, malware can threaten the security of computer systems at the hardware level by executing malicious applications \cite{b1}, \cite{b2}. Malware can be classified as Trojans, Viruses, and Worms. Traditional malware detection methods use antivirus software that is ineffective in detecting new malware. Since more than software-based malware detection is required due to the increasing number and variety of malware, hardware-based approaches have recently become a trend. In these approaches, it is assumed that while malicious attackers can modify software, compromising the hardware is challenging \cite{b3}.

While conventional software-based approaches for malware detections are inefficient in imposing complexity and computational overhead to systems, hardware-based approaches have been proposed by utilizing low-level micro-architectural features of running applications on the target system \cite{b1}. Hardware-based approaches employ Hardware Performance Counters (HPCs). HPCs are registers included in modern microprocessors, and their application monitors the system's performance. HPCs count microarchitectural low-level events such as instruction counts, hits, and misses in various cache levels and branches (mis)predictions during runtime. While there are more than 100 micro-architectural events for modern processors, the designed microprocessors have a limited number of HPC registers.  

For instance, on Intel, this limit number is four, and AMD has a limitation on six counter registers. This limitation in the number of HPC registers restricts monitoring HPCs in a small subset of events at one time \cite{b2}. Hardware Malware Detectors (HMDs) have been proposed to make computing systems more robust against malicious applications. HMDs can offer a significant advantage to defenders because the detector is always on and has little overhead for performance and power. While HMDs are assumed to be more robust against malware, the attackers can reverse engineer and compromise them. One possible attack could be changing the malware behavior and evading detection, which is a mimicry attack \cite{b5}. Evasive malware detection is not limited to the context of software malware detectors; it can also be a potential threat to HMDs where systematic reverse engineering is possible. \cite{b4}.

Recent developments and astonishing contributions of machine learning techniques for prediction problems with the significant size of data collected by high-performance computing systems and the Internet of Things became the motivation to use them to classify malware and benign applications by employing the large data set of HPCs. HPCs can count a wide variety of low-level events such as cache memories accesses and misses, TLB hits and misses, and branch mispredictions to optimize performance, energy-efficiency and improve the security of the system.

HPC implementation is a big challenge because significant numbers of micro-architectural events can be profiled to enhance the accuracy of monitoring low-level events. However, implementing hardware to support these vast amounts of features would not be cost-effective. A good trade-off between the accuracy and hardware implementation cost is required. For this purpose, ensemble learning approaches could be a game changer. They provide high-accuracy results compared to the general ML classifiers. After deploying feature reduction techniques, maintaining the accuracy of prediction and classification is essential. Hence, we utilized ensemble learning to achieve this goal.

Ensemble learning improves the accuracy and performance of general classifiers by taking advantage of a set of base learners and combining their outputs to make a final, accurate decision. This approach is efficient because it utilizes complementary information from different classifiers. There are various types of ensemble learning; in this work, we deploy and analyze two boosting and Bagging methods.

In this work, we evaluate binary malware classifiers and compare their performance detecting malware based on HPC monitoring while running malware. One of the most critical challenges in implementing hardware-based malware detection is to use the limited number of available HPC registers in a microprocessor (for instance, 2 or 4) without a drop in the accuracy and performance of malware detection. In this work, we make a comparison between using conventional general ML classifiers and ensemble learning to indicate the accuracy and performance of each of them. Among the most common ensemble learning methods, we choose one of the Boosting techniques, which are Adaptive Boosting, in short AdaBoost, and Bootstrap Aggregation, or in short, Bagging, which is a statistical prediction approach for classification and regression problems \cite{b4}. Therefore, in this experiment, we used 18 classifiers including tree-based classifiers such as Random Forest, Random Tree, J48, LMT, REPTREE, Hoeffding Tree, Decision Stump, function-based classifiers such as Multilayer Perceptron, SGD, SMO, Simple Logistic, and Logistic, rule-based classifiers such as JRIP, Decision Table, Part, and OneR, and the last groups are the Bayes Net and the Lazy classifier IBK. We examined and compared various HPCs for our project using 16, 8, 4, and 2 HPCs for original classifiers and 4 and 2 HPCs for Boosted and Bagging classifiers. Since adversarial attacks threaten the accuracy and performance of HMDs, it is crucial to consider their impact on the efficiency of HMDs. Hence, one of our future goals in this project is to explore the adversarial machine learning attacks of HMD and implement and evaluate them \cite{b9}  \cite{b72}.

\section{Related Work}

Malware is software designed to breach security and privacy. It's usually detected by antivirus software, which can be evaded by exploiting code vulnerabilities. Thus, hardware-based detection methods are becoming popular due to their robustness. Kanad Basu et al. \cite{b3} developed a mathematical model to assess malware detection probabilities using HPCs (hardware performance counters) monitored at set intervals. These counters, built into processors, track and optimize performance metrics \cite{p1}.

J. Demme et al. \cite{b6} examined the feasibility of building a malware detector in hardware using existing hardware performance counters. After reviewing a small set of variations on Android ARM and Intel Linux platforms, they aimed to detect malware and the variations within that family. Despite the existence of antivirus software, malware threats are growing, and there are several ways to work around the antivirus software. Though antivirus software decreases the danger of more malware, it has many shortcomings since antivirus software itself is prone to attack. Hence, the paper proposes a hardware method where hardware modifications are done to support the analysis of the program and detect malware by using performance counters. It collects micro-architectural traces of malware and applies machine learning algorithms such as KNN or decision trees to detect various malware with 90\% accuracy. It also describes the hardware support necessary to collect data and secure and update algorithms to prevent attacks. An essential assumption about malware that makes it possible for hardware detection is that all malware within a particular family, regardless of the code, attempts to do similar things. For example, if malware is used to collect GPS readings, that will not change even with code modification. As a result, it would go through a similar set of functional phases and exhibit similar detectable properties for the in-performance data that the authors exploit.

M. S. Islam et al. approach the question of whether malware can evade HMDs and if we can take steps to make the HMDs more resilient to this evasion. Several questions are reviewed and discussed in the existing literature. The paper shows how it is possible to accurately reverse engineer HMDs, regardless of their complexity. The author indicates that once HMDs are reverse-engineered, malware can easily evade it using low overhead evasion strategies. A new plan is developed in this paper — called timer interrupt injection, which allows the attacker to create powerful evasive malware while drastically reducing the overhead of the attack; more than 80\% and 60\% less dynamic overhead compared to the basic block, i.e., before every control flow altering instruction and function level, i.e., before every return instruction, injection strategies, respectively. The paper shows how complex classifiers like neural networks (NN) can adapt better to evasive malware than linear HMDs like LR. Despite that, new malware can easily reverse-engineer and evade such classifiers. A new class of resilient HMDs known as RHMDs has been developed and operates by randomizing detection across diverse detectors. The RHMD detection and evasion problem is formulated as a Bayesian Stackelberg game to establish an optimal switching strategy between the RHMD's base detectors that maximizes the detection accuracy of both evasive and non-evasive malware. Their results indicate that optimized RHMD switching can detect more than 45\% and 30\% of the malware and evasive malware missed by the RHMD detection, respectively \cite{b4}. Their threat model assumes that the attacker does not have white-box access to the RHMD and that they are unaware of the RHMD's precise implementation. If they did, the authors demonstrated that the attacker could theoretically elude detection at a cost proportional to the number of detectors.

In \cite{b2}, Zhou et al. proposed an experimental setup close to the user interface and evaluated the feasibility of machine learning models. Instead of relying on virtualization techniques, they run all the experiments in a natural environment. Further contributing to the realism of their experiments is the selection of training data for machine learning models. They test their models with measured HPC values from programs not observed during training. They claim that this represents a real-life scenario in which malware of the same category or family is available while learning machine learning models, but the user may encounter different malware. They identify common unrealistic assumptions and inadequate analysis that have been used in previous work using HPCs to detect malware; they train and test datasets similar to what previous works have done in the past, as well as in a realistic environment where the evaluated programs are not part of the training set. They compare the effects of this choice on the effectiveness of the ML models. 

Alam et al. \cite{b7} proposed their tools to learn the normal operation behavior of the system. HPC Event Time Series Data is given to an Artificial Neural Network to know specific data properties. Any deviation from this expected behavior learned by the automatic encoder is suspect to them. They were confronted with standard software that sounded the alarm, though benign in nature, which is a typical example of a benign process that differs significantly from the system's normal behavior due to high computational overhead, resulting in an incorrect warning. 

Anand et al. \cite{b71} presented a methodology for the early detection of ransomware using hardware performance counters (HPC). Recognizing the need for rapid detection due to ransomware's quick encryption capabilities, the authors focused on utilizing HPC, a feature often embedded in modern microprocessors, to monitor low-level system performance. 

Kuruvila et al. \cite {b8} demonstrated that the ready-made HPCs are not adequate to differentiate among individual applications due to their coarse granularity; they proposed their tailor-made HPCs, which capture the programs' fine granularity characteristics. It counts the sequence of instructions in the executed assembly code, and by mapping each instruction into five branch/jump(b), load(l), store(s), arithmetic(a), and Boolean (b) types, it tracks the sequence of instructions in the form of "XY" which increments each time the instruction of type "Y" is executed right after the instruction of type "X."

Karvandi et al. \cite{b5} presented that HyperDbg focuses on developing a new hardware-assisted debugger using hypervisor-level technology to provide high-performance and stealthy debugging capabilities. It is built on state-of-the-art hardware features like VT-x and Extended Page Table (EPT). It is designed to be OS-independent and avoids OS-based debugging APIs to enhance stealth against anti-debugging techniques.

Kadiyala et al. \cite{b51} demonstrated a fine-grained malware detection scheme by proposing a three-step methodology using HPCs. The steps include capturing HPCs at the system call level, dimensionality reduction, and machine learning classification. 

\section{Methodology}

\subsection{Overview of Experimental Platform}

%We set up our own Linux Container using Docker and applied perf utility on the Linux Kernel under privileged access. 

%The training and testing dataset used in this project consisted of a feature reduced dataset of HPCs with features like branch instructions, L1 cache misses and so on. The features include bus cycles (time to perform a read/write between CPU and memory), branch instructions (branch instructions retired), branch misses (branch miss predicted), branch loads (successful branches), cache references (last level cache reference), cache misses (last level cache misses), node-loads (successful load operations to DRAM), node stores (successful store operations to DRAM), instructions (instructions retired), layer1 data cache loads (retired memory load operations), layer 1 data cache stores, (cache lines stored into layer 1 cache from DRAM), layer 1 instruction cache load misses (instruction misses in layer 1 instruction cache), llc\_load\_miss (cache lines brought into layer 3 cache from DRAM), llc\_loads  (successful memory load operations in layer 3 cache), itlb-loads- misses (misses in instruction TLB during load operations).

We established a Linux Container through Docker and utilized the Perf tool on the Linux Kernel with privileged access. The training and testing data comprised a dataset of HPCs that includes 16 different hardware counter measurements. The HPC events and their descriptions are available in Table I.

%This dataset was not a balanced set, having almost double the number of benign entries as compared to malware entries. After reviewing the Evaluation metrics for unbalanced dataset – which was clearly biased towards benign applications, we decided to use a balanced dataset – new dataset consists of 3700 samples of both benign and malware dataset entries. The dataset was split between 70\% for training set and 30\% for testing set in the Weka Tool. The idea of the project is to collect a dataset using perf tools on a Linux Kernel and further use that dataset to train and analyze the machine learning classifiers. Since the dataset we’re dealing with here consists of the profiling of several commands and classifying them into benign and malware categories, multiple malwares as being dealt with. For this use case, a Linux container - Docker is the best option.

The dataset initially was imbalanced, with benign entries nearly twice the number of malware entries. Upon evaluating the metrics for the imbalanced dataset, which favored benign applications, we opted for a balanced dataset comprising 3700 samples each of benign and malware entries. We divided the dataset into a 70\% training set and a 30\% testing set. The project's goal is to gather a dataset through perf tools on a Linux Kernel, then employ this dataset to train and evaluate machine learning classifiers. Given that the dataset involves profiling various commands to distinguish between benign and malware categories, handling multiple malware becomes necessary. In this case, the ideal solution is to use a Linux container, such as Docker.

Linux performance counters are a mature subsystem in the kernel, intended for performance analysis. These include hardware features like CPU/PMU (Performance Monitoring Unit) and software elements like software counters and trace points. The Perf tool specific to Linux uses a system call named perf events to interface with the kernel and collect the HPC values needed for profiling activities.

Docker is a technology that allows developers, system administrators and IT professionals to run applications in isolated environments called containers on host operating systems such as Linux. A key advantage of Docker is its ability to encapsulate an application along with its required dependencies into a consistent unit for software deployment. Containers provide a more resource-efficient alternative to virtual machines and facilitate a secure environment to execute various malware analyses without risking the system's security.

%\begin{figure}[htbp]%\centerline{\includegraphics[width=\columnwidth]{./Figures/Figure1.eps}} %\caption{Docker Container Architecture \cite{b11}}\label{fig}%\end{figure} \subsection{Dataset} Table I shows the attributes in the dataset. Bus-cycles are a measure of the number of CPU cycles spent reading/writing to/from memory. Branch-instructions counts the number of branch instructions that have been processed and finished. Branch-misses counts how many times the target of a branch instruction is wrongly predicted by the CPU. The branch-loads are the number of successfully executed branch instructions. Cache-references is the number of accesses to the last level cache, and Cache-misses is the number of requests for data by the CPU that was not in the last level cache. Node-loads count successful operations that retrieve data from DRAM and Node-stores count successful operations that store data to DRAM. The instructions field specifies the total number of instructions executed by the CPU. L1-dcache-loads are memory load operations completed by the L1 data cache. (L1-dcache-stores count cache line stores from DRAM into the L1 data cache.) L1-icache-load-misses happen when the CPU can’t find the instructions in the L1 instruction cache. The L1-dcache-load-misses measures the number of L1 data cache load misses. LLC-load-misses are the misses in the L3 cache in loading data from DRAM. LLC-loads are the successful load operations in the L3 cache. Finally, iTLB-load-misses count the number of misses in the Instruction Translation Lookaside Buffer during load operations.

\subsection{Dataset}
The dataset contains the attributes shown in Table I. Bus-cycles measure the number of CPU cycles spent on read/write operations between the CPU and memory. Branch-instructions count the number of branch instructions that have been issued and completed, and Branch-misses count how often the CPU has incorrectly predicted the target of a branch instruction. Branch-loads are the number of successfully executed branch instructions. Cache-references are the number of times the last level cache has been accessed, and Cache-misses are the number of times the CPU has not found the requested data in the last level cache. Node-loads are the number of successful data retrievals from DRAM, and Node-stores are the number of successful data storage operations to DRAM. Instructions are the total number of instructions completed by the CPU. L1-dcache-loads are memory load operations completed by the L1 data cache, and L1-dcache-stores are the number of cache lines stored from DRAM to the L1 data cache. L1-icache-load-misses are the number of times the CPU has not found the instructions in the L1 instruction cache. L1-dcache-load-misses are the number of misses in the L1 data cache in the course of load operations. LLC-load-misses are the number of misses in the L3 cache while trying to load data from DRAM, and LLC-loads are the number of successful load operations in the L3 cache. iTLB-load-misses are the number of misses in the Instruction Translation Lookaside Buffer in the course of load operations.

In our analysis, as shown in Table II, we analyzed the correlation among various attributes to identify relevant features for machine learning (ML) classifiers. Based on this analysis, five feature islands were identified and subsequently evaluated using a Ranker algorithm to determine their significance in training ML classifiers. To simplify the process, we reduced the original feature set to three subsets: HPC 8, HPC 4, and HPC 2 datasets. These subsets were then analyzed separately to assess the performance of the ML classifiers.

The results of the feature reduction process are summarized in Table III, which shows the features that have been preserved in the 8, 4 and 2 HPC modes with their respective rankings based on their contribution to the effectiveness of the classifiers. For instance, the features 'L1-dcache-loads' and 'L1-dcache-stores' are shown to be highly significant across the three datasets by virtue of their inclusion and rankings. Features such as 'Branch-instructions' and 'Instructions' were only seen in the 8 HPC dataset, implying their less relevance in smaller feature sets. This table provides a good summary of the most influential features in different circumstances, and thus offers one to better configure ML classifiers for better performance.

%TTT Table1
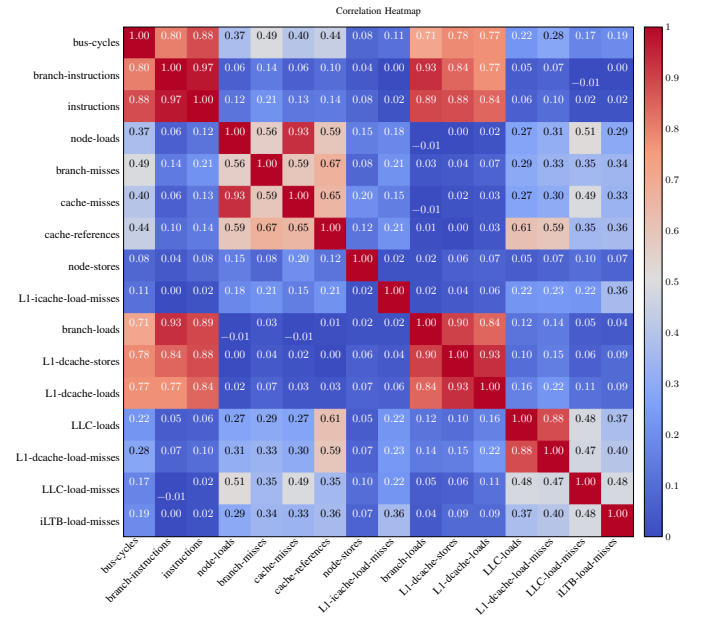
\begin{table}[htbp]
\centering
\caption{HPC features with their descriptions}
\begin{tabularx}{\columnwidth}{lX} % Notice the use of 'X' for the second column to allow text wrapping
\toprule
\textbf{HPC Event} & \textbf{Description} \\
\midrule
Bus-cycles & Time to do an R/W between CPU and memory \\
Branch-instructions & Branch instructions retired \\
Branch-misses & Branch mis-predicted \\
Branch-loads & Successful branches \\
Cache-references & Last level cache references \\
Cache-misses & Last level cache misses \\
Node-loads & Successful load operations to DRAM \\
Node-stores & Successful store operations to DRAM \\
Instructions & Instructions retired \\
L1-dcache-loads & Retired memory load operations \\
L1-dcache-stores & Cache lines stored into L1 cache from DRAM \\
L1-icache-load-misses & Instruction misses in L1 instruction cache \\
L1-dcache-load-misses & Misses in cache lines into L1 dcache \\
LLC-load-misses & Cache lines brought into L3 cache from DRAM \\
LLC-loads & Successful memory load operations in L3 cache \\
iTLB-load-misses & Misses in  ITLB during load operations \\
\bottomrule
\end{tabularx}
\end{table}

%TTT Table2
\begin{table}[htbp]
\centering
\caption{Correlation Heatmap}
\resizebox{\columnwidth}{!}{%
\begin{tikzpicture}
\begin{axis}[
    title={Correlation Heatmap},
    width=16cm, height=16cm,
    enlargelimits=false,
    axis on top,
    xmin=-0.5, xmax=15.5,
    ymin=-0.5, ymax=15.5,
    y dir=reverse,
    xtick={0,...,15}, ytick={0,...,15},
    xticklabels={bus-cycles, branch-instructions, instructions, node-loads,
                 branch-misses, cache-misses, cache-references, node-stores,
                 L1-icache-load-misses, branch-loads, L1-dcache-stores,
                 L1-dcache-loads, LLC-loads, L1-dcache-load-misses,
                 LLC-load-misses, iLTB-load-misses},
    yticklabels={bus-cycles, branch-instructions, instructions, node-loads,
                 branch-misses, cache-misses, cache-references, node-stores,
                 L1-icache-load-misses, branch-loads, L1-dcache-stores,
                 L1-dcache-loads, LLC-loads, L1-dcache-load-misses,
                 LLC-load-misses, iLTB-load-misses},
    x tick label style={rotate=45, anchor=east, font=\small},
    y tick label style={font=\small},
    tick style={draw=none},
    colorbar,
    colormap name=coolwarm,
    point meta min=0, point meta max=1,
    nodes near coords={\pgfmathprintnumber[fixed,fixed zerofill,precision=2]{\pgfplotspointmeta}},
    every node near coord/.append style={
        font=\small,
        /utils/exec={\pgfmathfloattofixed{\pgfplotspointmeta}%
                     \let\cellval\pgfmathresult},
        text=\ifdim\cellval pt<0.25pt white\else
             \ifdim\cellval pt>0.70pt white\else black\fi\fi,
    },
]
\addplot[matrix plot*, mesh/cols=16, point meta=explicit] coordinates {
(0,0)[1.00] (1,0)[0.80] (2,0)[0.88] (3,0)[0.37] (4,0)[0.49] (5,0)[0.40] (6,0)[0.44] (7,0)[0.08] (8,0)[0.11] (9,0)[0.71] (10,0)[0.78] (11,0)[0.77] (12,0)[0.22] (13,0)[0.28] (14,0)[0.17] (15,0)[0.19]

(0,1)[0.80] (1,1)[1.00] (2,1)[0.97] (3,1)[0.06] (4,1)[0.14] (5,1)[0.06] (6,1)[0.10] (7,1)[0.04] (8,1)[0.00] (9,1)[0.93] (10,1)[0.84] (11,1)[0.77] (12,1)[0.05] (13,1)[0.07] (14,1)[-0.01] (15,1)[0.00]

(0,2)[0.88] (1,2)[0.97] (2,2)[1.00] (3,2)[0.12] (4,2)[0.21] (5,2)[0.13] (6,2)[0.14] (7,2)[0.08] (8,2)[0.02] (9,2)[0.89] (10,2)[0.88] (11,2)[0.84] (12,2)[0.06] (13,2)[0.10] (14,2)[0.02] (15,2)[0.02]

(0,3)[0.37] (1,3)[0.06] (2,3)[0.12] (3,3)[1.00] (4,3)[0.56] (5,3)[0.93] (6,3)[0.59] (7,3)[0.15] (8,3)[0.18] (9,3)[-0.01] (10,3)[0.00] (11,3)[0.02] (12,3)[0.27] (13,3)[0.31] (14,3)[0.51] (15,3)[0.29]

(0,4)[0.49] (1,4)[0.14] (2,4)[0.21] (3,4)[0.56] (4,4)[1.00] (5,4)[0.59] (6,4)[0.67] (7,4)[0.08] (8,4)[0.21] (9,4)[0.03] (10,4)[0.04] (11,4)[0.07] (12,4)[0.29] (13,4)[0.33] (14,4)[0.35] (15,4)[0.34]

(0,5)[0.40] (1,5)[0.06] (2,5)[0.13] (3,5)[0.93] (4,5)[0.59] (5,5)[1.00] (6,5)[0.65] (7,5)[0.20] (8,5)[0.15] (9,5)[-0.01] (10,5)[0.02] (11,5)[0.03] (12,5)[0.27] (13,5)[0.30] (14,5)[0.49] (15,5)[0.33]

(0,6)[0.44] (1,6)[0.10] (2,6)[0.14] (3,6)[0.59] (4,6)[0.67] (5,6)[0.65] (6,6)[1.00] (7,6)[0.12] (8,6)[0.21] (9,6)[0.01] (10,6)[0.00] (11,6)[0.03] (12,6)[0.61] (13,6)[0.59] (14,6)[0.35] (15,6)[0.36]

(0,7)[0.08] (1,7)[0.04] (2,7)[0.08] (3,7)[0.15] (4,7)[0.08] (5,7)[0.20] (6,7)[0.12] (7,7)[1.00] (8,7)[0.02] (9,7)[0.02] (10,7)[0.06] (11,7)[0.07] (12,7)[0.05] (13,7)[0.07] (14,7)[0.10] (15,7)[0.07]

(0,8)[0.11] (1,8)[0.00] (2,8)[0.02] (3,8)[0.18] (4,8)[0.21] (5,8)[0.15] (6,8)[0.21] (7,8)[0.02] (8,8)[1.00] (9,8)[0.02] (10,8)[0.04] (11,8)[0.06] (12,8)[0.22] (13,8)[0.23] (14,8)[0.22] (15,8)[0.36]

(0,9)[0.71] (1,9)[0.93] (2,9)[0.89] (3,9)[-0.01] (4,9)[0.03] (5,9)[-0.01] (6,9)[0.01] (7,9)[0.02] (8,9)[0.02] (9,9)[1.00] (10,9)[0.90] (11,9)[0.84] (12,9)[0.12] (13,9)[0.14] (14,9)[0.05] (15,9)[0.04]

(0,10)[0.78] (1,10)[0.84] (2,10)[0.88] (3,10)[0.00] (4,10)[0.04] (5,10)[0.02] (6,10)[0.00] (7,10)[0.06] (8,10)[0.04] (9,10)[0.90] (10,10)[1.00] (11,10)[0.93] (12,10)[0.10] (13,10)[0.15] (14,10)[0.06] (15,10)[0.09]

(0,11)[0.77] (1,11)[0.77] (2,11)[0.84] (3,11)[0.02] (4,11)[0.07] (5,11)[0.03] (6,11)[0.03] (7,11)[0.07] (8,11)[0.06] (9,11)[0.84] (10,11)[0.93] (11,11)[1.00] (12,11)[0.16] (13,11)[0.22] (14,11)[0.11] (15,11)[0.09]

(0,12)[0.22] (1,12)[0.05] (2,12)[0.06] (3,12)[0.27] (4,12)[0.29] (5,12)[0.27] (6,12)[0.61] (7,12)[0.05] (8,12)[0.22] (9,12)[0.12] (10,12)[0.10] (11,12)[0.16] (12,12)[1.00] (13,12)[0.88] (14,12)[0.48] (15,12)[0.37]

(0,13)[0.28] (1,13)[0.07] (2,13)[0.10] (3,13)[0.31] (4,13)[0.33] (5,13)[0.30] (6,13)[0.59] (7,13)[0.07] (8,13)[0.23] (9,13)[0.14] (10,13)[0.15] (11,13)[0.22] (12,13)[0.88] (13,13)[1.00] (14,13)[0.47] (15,13)[0.40]

(0,14)[0.17] (1,14)[-0.01] (2,14)[0.02] (3,14)[0.51] (4,14)[0.35] (5,14)[0.49] (6,14)[0.35] (7,14)[0.10] (8,14)[0.22] (9,14)[0.05] (10,14)[0.06] (11,14)[0.11] (12,14)[0.48] (13,14)[0.47] (14,14)[1.00] (15,14)[0.48]

(0,15)[0.19] (1,15)[0.00] (2,15)[0.02] (3,15)[0.29] (4,15)[0.34] (5,15)[0.33] (6,15)[0.36] (7,15)[0.07] (8,15)[0.36] (9,15)[0.04] (10,15)[0.09] (11,15)[0.09] (12,15)[0.37] (13,15)[0.40] (14,15)[0.48] (15,15)[1.00]
};
\end{axis}
\end{tikzpicture}%
}
\end{table}

%TTT Table3
\begin{table}[htbp]
\centering
\caption{Reduced Features within 8, 4, and 2 HPC modes}
\begin{tabularx}{\columnwidth}{@{}Xcccc@{}}
\toprule
\textbf{Classifier} & \textbf{Ranked} & \textbf{8 HPC} & \textbf{4 HPC} & \textbf{2 HPC} \\ \midrule
L1-dcache-loads & 0.516 & $\checkmark$ & $\checkmark$ & $\checkmark$ \\
L1-dcache-stores & 0.492 & $\checkmark$ & $\checkmark$ & $\checkmark$ \\
Branch-loads & 0.455 & $\checkmark$ & $\checkmark$ &  \\
Branch-misses & 0.452 & $\checkmark$ & $\checkmark$ &  \\
Branch-instructions & 0.449 & $\checkmark$ & &  \\
Instructions & 0.444 & $\checkmark$ & &  \\
Bus-cycles & 0.321 & $\checkmark$ & &  \\
L1-icache-load-misses & 0.278 & $\checkmark$ & &  \\ \bottomrule
\end{tabularx}
\label{your-label-here}
\end{table}

\section{Evaluation Framework and Experimental Results}

In this section, we conduct a comprehensive evaluation of 144 different experiments with classifiers for malware detection, with a focus on their accuracy, area under the ROC curve, and overall performance. Ultimately, we evaluate every classifier according to its performance metric to offer suggestions regarding the design of hardware-based malware detection systems.

\subsection{Evaluation Framework}

Accuracy shows the percentage of predictions that were right and is defined as the ratio of the number of correct predictions to the total number of predictions. It is defined as:

\begin{equation}
\text{Accuracy (\%)} = \frac{TP + TN}{TP + TN + FP + FN} \times 100
\end{equation}

 which calculates the proportion of correct predictions (both true positives and true negatives) out of all predictions made (the total of true positives, true negatives, false positives, and false negatives). The results are shown in Table IV.

% \usepackage{multirow}
\begin{table}[]
\centering
\caption{A Comparison of classifiers based on the Accuracy}
\resizebox{\columnwidth}{!}{%
\begin{tabular}{|c|ccccc|ccc|}
\hline
\rowcolor[HTML]{D5A6BD} 
\cellcolor[HTML]{D5A6BD}                             & \multicolumn{1}{c|}{\cellcolor[HTML]{D5A6BD}HPC16} & \multicolumn{1}{c|}{\cellcolor[HTML]{D5A6BD}HPC8}  & \multicolumn{3}{c|}{\cellcolor[HTML]{D5A6BD}HPC4}                                                                                         & \multicolumn{3}{c|}{\cellcolor[HTML]{D5A6BD}HPC2}                                                                                          \\ \cline{2-9} 
\rowcolor[HTML]{D5A6BD} 
\multirow{-2}{*}{\cellcolor[HTML]{D5A6BD}Classifier} & \multicolumn{3}{c|}{\cellcolor[HTML]{D5A6BD}Single}                                                                                                          & \multicolumn{1}{c|}{\cellcolor[HTML]{D5A6BD}Boosted} & Bagging                       & \multicolumn{1}{c|}{\cellcolor[HTML]{D5A6BD}Single} & \multicolumn{1}{c|}{\cellcolor[HTML]{D5A6BD}Boosted} & Bagging                       \\ \hline
\cellcolor[HTML]{9FC5E8}MLP                          & \multicolumn{1}{c|}{\cellcolor[HTML]{FCC27C}82.34} & \multicolumn{1}{c|}{\cellcolor[HTML]{FCC07B}81.98} & \multicolumn{1}{c|}{\cellcolor[HTML]{FBA877}78.96} & \multicolumn{1}{c|}{\cellcolor[HTML]{FBB178}80.14}   & \cellcolor[HTML]{FBA977}79.01 & \multicolumn{1}{c|}{\cellcolor[HTML]{FA9673}76.58}  & \multicolumn{1}{c|}{\cellcolor[HTML]{FA9673}76.58}   & \cellcolor[HTML]{FA9673}76.58 \\ \hline
\cellcolor[HTML]{9FC5E8}SGD                          & \multicolumn{1}{c|}{\cellcolor[HTML]{FBB078}79.96} & \multicolumn{1}{c|}{\cellcolor[HTML]{FBA576}78.51} & \multicolumn{1}{c|}{\cellcolor[HTML]{FA9C74}77.39} & \multicolumn{1}{c|}{\cellcolor[HTML]{FA9D75}77.57}   & \cellcolor[HTML]{FA9C74}77.43 & \multicolumn{1}{c|}{\cellcolor[HTML]{F8756D}72.43}  & \multicolumn{1}{c|}{\cellcolor[HTML]{F86E6C}71.53}   & \cellcolor[HTML]{F8766D}72.57 \\ \hline
\cellcolor[HTML]{9FC5E8}SimpleLogistic               & \multicolumn{1}{c|}{\cellcolor[HTML]{FBAC77}79.46} & \multicolumn{1}{c|}{\cellcolor[HTML]{FA9E75}77.61} & \multicolumn{1}{c|}{\cellcolor[HTML]{FA9273}76.17} & \multicolumn{1}{c|}{\cellcolor[HTML]{FA9273}76.17}   & \cellcolor[HTML]{FA9272}76.08 & \multicolumn{1}{c|}{\cellcolor[HTML]{F86E6C}71.53}  & \multicolumn{1}{c|}{\cellcolor[HTML]{F86E6C}71.53}   & \cellcolor[HTML]{F86E6C}71.44 \\ \hline
\cellcolor[HTML]{9FC5E8}SMO                          & \multicolumn{1}{c|}{\cellcolor[HTML]{FBAE78}79.68} & \multicolumn{1}{c|}{\cellcolor[HTML]{FAA075}77.93} & \multicolumn{1}{c|}{\cellcolor[HTML]{FA9B74}77.30} & \multicolumn{1}{c|}{\cellcolor[HTML]{FA9E75}77.66}   & \cellcolor[HTML]{FA9B74}77.25 & \multicolumn{1}{c|}{\cellcolor[HTML]{F8776D}72.61}  & \multicolumn{1}{c|}{\cellcolor[HTML]{F8776D}72.61}   & \cellcolor[HTML]{F8776D}72.61 \\ \hline
\cellcolor[HTML]{9FC5E8}Logistic                     & \multicolumn{1}{c|}{\cellcolor[HTML]{FBAC77}79.41} & \multicolumn{1}{c|}{\cellcolor[HTML]{FBA075}77.97} & \multicolumn{1}{c|}{\cellcolor[HTML]{FA9273}76.17} & \multicolumn{1}{c|}{\cellcolor[HTML]{FA9273}76.17}   & \cellcolor[HTML]{FA9373}76.22 & \multicolumn{1}{c|}{\cellcolor[HTML]{F8696B}70.77}  & \multicolumn{1}{c|}{\cellcolor[HTML]{F8696B}70.77}   & \cellcolor[HTML]{F8696B}70.77 \\ \hline
\cellcolor[HTML]{B6D7A8}RandomForest                 & \multicolumn{1}{c|}{\cellcolor[HTML]{63BE7B}98.33} & \multicolumn{1}{c|}{\cellcolor[HTML]{75C47D}97.12} & \multicolumn{1}{c|}{\cellcolor[HTML]{8ACA7E}95.63} & \multicolumn{1}{c|}{\cellcolor[HTML]{8BCA7E}95.59}   & \cellcolor[HTML]{8BCA7E}95.59 & \multicolumn{1}{c|}{\cellcolor[HTML]{E7E483}89.19}  & \multicolumn{1}{c|}{\cellcolor[HTML]{E9E583}89.05}   & \cellcolor[HTML]{EBE683}88.92 \\ \hline
\cellcolor[HTML]{B6D7A8}RandomTree                   & \multicolumn{1}{c|}{\cellcolor[HTML]{8ACA7E}95.68} & \multicolumn{1}{c|}{\cellcolor[HTML]{95CD7E}94.86} & \multicolumn{1}{c|}{\cellcolor[HTML]{A0D07F}94.14} & \multicolumn{1}{c|}{\cellcolor[HTML]{A2D17F}93.96}   & \cellcolor[HTML]{86C87D}95.95 & \multicolumn{1}{c|}{\cellcolor[HTML]{FDEB84}87.70}  & \multicolumn{1}{c|}{\cellcolor[HTML]{FDEB84}87.70}   & \cellcolor[HTML]{F2E784}88.47 \\ \hline
\cellcolor[HTML]{B6D7A8}J48                          & \multicolumn{1}{c|}{\cellcolor[HTML]{7DC67D}96.58} & \multicolumn{1}{c|}{\cellcolor[HTML]{83C87D}96.13} & \multicolumn{1}{c|}{\cellcolor[HTML]{A4D17F}93.87} & \multicolumn{1}{c|}{\cellcolor[HTML]{8ACA7E}95.68}   & \cellcolor[HTML]{93CC7E}95.00 & \multicolumn{1}{c|}{\cellcolor[HTML]{F9EA84}87.93}  & \multicolumn{1}{c|}{\cellcolor[HTML]{FFEB84}87.57}   & \cellcolor[HTML]{E7E483}89.19 \\ \hline
\cellcolor[HTML]{B6D7A8}LMT                          & \multicolumn{1}{c|}{\cellcolor[HTML]{7AC57D}96.76} & \multicolumn{1}{c|}{\cellcolor[HTML]{88C97E}95.77} & \multicolumn{1}{c|}{\cellcolor[HTML]{9BCF7F}94.46} & \multicolumn{1}{c|}{\cellcolor[HTML]{8CCA7E}95.50}   & \cellcolor[HTML]{92CC7E}95.09 & \multicolumn{1}{c|}{\cellcolor[HTML]{E7E483}89.19}  & \multicolumn{1}{c|}{\cellcolor[HTML]{F8E984}88.02}   & \cellcolor[HTML]{ECE683}88.87 \\ \hline
\cellcolor[HTML]{B6D7A8}REPTree                      & \multicolumn{1}{c|}{\cellcolor[HTML]{85C87D}95.99} & \multicolumn{1}{c|}{\cellcolor[HTML]{97CD7E}94.73} & \multicolumn{1}{c|}{\cellcolor[HTML]{A4D17F}93.83} & \multicolumn{1}{c|}{\cellcolor[HTML]{8FCB7E}95.32}   & \cellcolor[HTML]{9CCF7F}94.41 & \multicolumn{1}{c|}{\cellcolor[HTML]{ECE683}88.83}  & \multicolumn{1}{c|}{\cellcolor[HTML]{EEE784}88.69}   & \cellcolor[HTML]{EAE583}88.96 \\ \hline
\cellcolor[HTML]{B6D7A8}HoeffdingTree                & \multicolumn{1}{c|}{\cellcolor[HTML]{FBAA77}79.19} & \multicolumn{1}{c|}{\cellcolor[HTML]{FA9974}76.98} & \multicolumn{1}{c|}{\cellcolor[HTML]{FA9974}76.98} & \multicolumn{1}{c|}{\cellcolor[HTML]{FA9473}76.40}   & \cellcolor[HTML]{FAA075}77.93 & \multicolumn{1}{c|}{\cellcolor[HTML]{FA9172}76.04}  & \multicolumn{1}{c|}{\cellcolor[HTML]{FA9072}75.90}   & \cellcolor[HTML]{FA9172}76.04 \\ \hline
\rowcolor[HTML]{FA9172} 
\cellcolor[HTML]{B6D7A8}DecisionStump                & \multicolumn{1}{c|}{\cellcolor[HTML]{FA9172}76.04} & \multicolumn{1}{c|}{\cellcolor[HTML]{FA9172}76.04} & \multicolumn{1}{c|}{\cellcolor[HTML]{FA9172}76.04} & \multicolumn{1}{c|}{\cellcolor[HTML]{FDC77D}82.93}   & 76.04                         & \multicolumn{1}{c|}{\cellcolor[HTML]{FA9172}76.04}  & \multicolumn{1}{c|}{\cellcolor[HTML]{FA9172}76.04}   & 76.04                         \\ \hline
\cellcolor[HTML]{B4A7D6}JRIP                         & \multicolumn{1}{c|}{\cellcolor[HTML]{82C77D}96.22} & \multicolumn{1}{c|}{\cellcolor[HTML]{A0D07F}94.10} & \multicolumn{1}{c|}{\cellcolor[HTML]{B2D580}92.88} & \multicolumn{1}{c|}{\cellcolor[HTML]{A5D17F}93.78}   & \cellcolor[HTML]{9BCE7F}94.50 & \multicolumn{1}{c|}{\cellcolor[HTML]{FEE983}87.25}  & \multicolumn{1}{c|}{\cellcolor[HTML]{F9EA84}87.93}   & \cellcolor[HTML]{F4E884}88.33 \\ \hline
\cellcolor[HTML]{B4A7D6}DecisionTable                & \multicolumn{1}{c|}{\cellcolor[HTML]{8ACA7E}95.63} & \multicolumn{1}{c|}{\cellcolor[HTML]{A4D17F}93.87} & \multicolumn{1}{c|}{\cellcolor[HTML]{BDD881}92.12} & \multicolumn{1}{c|}{\cellcolor[HTML]{A2D17F}93.96}   & \cellcolor[HTML]{C2DA81}91.80 & \multicolumn{1}{c|}{\cellcolor[HTML]{F9EA84}87.93}  & \multicolumn{1}{c|}{\cellcolor[HTML]{F6E984}88.15}   & \cellcolor[HTML]{FBEA84}87.79 \\ \hline
\cellcolor[HTML]{B4A7D6}Part                         & \multicolumn{1}{c|}{\cellcolor[HTML]{80C77D}96.35} & \multicolumn{1}{c|}{\cellcolor[HTML]{AAD380}93.42} & \multicolumn{1}{c|}{\cellcolor[HTML]{B8D780}92.48} & \multicolumn{1}{c|}{\cellcolor[HTML]{8CCA7E}95.50}   & \cellcolor[HTML]{A5D17F}93.78 & \multicolumn{1}{c|}{\cellcolor[HTML]{FEE382}86.53}  & \multicolumn{1}{c|}{\cellcolor[HTML]{FAEA84}87.88}   & \cellcolor[HTML]{F1E784}88.51 \\ \hline
\cellcolor[HTML]{B4A7D6}OneR                         & \multicolumn{1}{c|}{\cellcolor[HTML]{FDCE7E}83.83} & \multicolumn{1}{c|}{\cellcolor[HTML]{FDCE7E}83.83} & \multicolumn{1}{c|}{\cellcolor[HTML]{FDCE7E}83.83} & \multicolumn{1}{c|}{\cellcolor[HTML]{FEDA80}85.32}   & \cellcolor[HTML]{FDD57F}84.68 & \multicolumn{1}{c|}{\cellcolor[HTML]{FDCE7E}83.83}  & \multicolumn{1}{c|}{\cellcolor[HTML]{FDC97D}83.20}   & \cellcolor[HTML]{FDD57F}84.68 \\ \hline
\cellcolor[HTML]{FFE599}BayesNet                     & \multicolumn{1}{c|}{\cellcolor[HTML]{82C77D}96.22} & \multicolumn{1}{c|}{\cellcolor[HTML]{E0E283}89.68} & \multicolumn{1}{c|}{\cellcolor[HTML]{FEEA83}87.43} & \multicolumn{1}{c|}{\cellcolor[HTML]{FEE683}86.98}   & \cellcolor[HTML]{FEE783}87.07 & \multicolumn{1}{c|}{\cellcolor[HTML]{FDD780}84.96}  & \multicolumn{1}{c|}{\cellcolor[HTML]{FDD37F}84.41}   & \cellcolor[HTML]{FDD67F}84.82 \\ \hline
\cellcolor[HTML]{F9CB9C}KNN                          & \multicolumn{1}{c|}{\cellcolor[HTML]{71C27C}97.39} & \multicolumn{1}{c|}{\cellcolor[HTML]{89C97E}95.72} & \multicolumn{1}{c|}{\cellcolor[HTML]{98CE7F}94.68} & \multicolumn{1}{c|}{\cellcolor[HTML]{98CE7F}94.68}   & \cellcolor[HTML]{99CE7F}94.64 & \multicolumn{1}{c|}{\cellcolor[HTML]{E8E583}89.14}  & \multicolumn{1}{c|}{\cellcolor[HTML]{E8E583}89.14}   & \cellcolor[HTML]{EEE683}88.74 \\ \hline
\end{tabular}}
\end{table}

%%FFF Fig2
%\begin{figure}[htbp]
%\centerline{\includegraphics[width=\columnwidth]%{Figures/Figure2.png}}
%\caption{Accuracy of tree-based classifiers}
%\label{fig}
%\end{figure}

%%FFF Fig3
%\begin{figure}[htbp]
%\centerline{\includegraphics[width=\columnwidth]%{Figures/Figure3.png}}
%\caption{Accuracy of rule-based classifiers}
%\label{fig}
%\end{figure}

%%FFF Fig4
%\begin{figure}[htbp]
 %   \centerline{\includegraphics[width=\columnwidth]{Figures/Figure4.png}}
%\caption{Accuracy of Bayes and Lazy classifiers}
%\label{fig}
%\end{figure}

%\subsection{Precision Metric}

%Precision shows the percentage of positive identifications that were classified correct. It is defined as:

%\begin{equation}
%\text{Precision (\%)} = \left( \frac{TP}{TP + FP} \right) \times 100
%\end{equation}

%\subsection{Recall Metric}

%Recall shows the percentage of actual positives that were identified correctly. It is defined as:

%\begin{equation}
%\text{Recall (\%)} = \left( \frac{TP}{TP + FN} \right) \times 100
%\end{equation}

%\subsection{F Measure Metric}

%$F_1$ score takes both precision and recall into account and is defined as the harmonic mean of precision and recall metrics.

%\begin{equation}
%F1 = 2 \times \frac{\text{Precision} \times \text{Recall}}{\text{Precision} + \text{Recall}}
%\end{equation}

The Area Under ROC Curve metric is related to the robustness of an ML classifier; it is a number between 0.5 and 1.0, with 1.0 illustrating better robustness and 0.5 standing for a random guess. The higher value of AUC shows the ROC curve is closer to the ideal classifier, and it is performing better in classification of the benign and malicious applications. The Area Under the ROC Curve is calculated as:

\begin{equation}
\text{AUC(ROC)} = \int \text{TPR}(\text{FPR}) \, d(\text{FPR}) 
\end{equation}

where the TPR is integrated over the range of possible FPR values. This integration effectively sums up the performance of the classifier at all possible classification thresholds. Table V shows the result of area under ROC for the trained classifiers. 

% \usepackage{multirow}
\begin{table}[]
\centering
\caption{A Comparison of classifiers based on the Area Under ROC Curve}
\resizebox{\columnwidth}{!}{%
\begin{tabular}{|c|ccccc|ccc|}
\hline
\rowcolor[HTML]{D5A6BD} 
\cellcolor[HTML]{D5A6BD}                             & \multicolumn{1}{c|}{\cellcolor[HTML]{D5A6BD}HPC16} & \multicolumn{1}{c|}{\cellcolor[HTML]{D5A6BD}HPC8}  & \multicolumn{3}{c|}{\cellcolor[HTML]{D5A6BD}HPC4}                                                                                         & \multicolumn{3}{c|}{\cellcolor[HTML]{D5A6BD}HPC2}                                                                                          \\ \cline{2-9} 
\rowcolor[HTML]{D5A6BD} 
\multirow{-2}{*}{\cellcolor[HTML]{D5A6BD}Classifier} & \multicolumn{3}{c|}{\cellcolor[HTML]{D5A6BD}Single}                                                                                                          & \multicolumn{1}{c|}{\cellcolor[HTML]{D5A6BD}Boosted} & Bagging                       & \multicolumn{1}{c|}{\cellcolor[HTML]{D5A6BD}Single} & \multicolumn{1}{c|}{\cellcolor[HTML]{D5A6BD}Boosted} & Bagging                       \\ \hline
\cellcolor[HTML]{9FC5E8}MLP                          & \multicolumn{1}{c|}{\cellcolor[HTML]{C1DA81}0.96}  & \multicolumn{1}{c|}{\cellcolor[HTML]{FDD27F}0.895} & \multicolumn{1}{c|}{\cellcolor[HTML]{FCB77A}0.851} & \multicolumn{1}{c|}{\cellcolor[HTML]{FDCC7E}0.885}   & \cellcolor[HTML]{FED880}0.905 & \multicolumn{1}{c|}{\cellcolor[HTML]{FA9172}0.79}   & \multicolumn{1}{c|}{\cellcolor[HTML]{FA9172}0.79}    & \cellcolor[HTML]{FA9C74}0.808 \\ \hline
\cellcolor[HTML]{9FC5E8}SGD                          & \multicolumn{1}{c|}{\cellcolor[HTML]{FA9874}0.802} & \multicolumn{1}{c|}{\cellcolor[HTML]{FA8F72}0.787} & \multicolumn{1}{c|}{\cellcolor[HTML]{F98871}0.776} & \multicolumn{1}{c|}{\cellcolor[HTML]{FBA476}0.822}   & \cellcolor[HTML]{F98A71}0.779 & \multicolumn{1}{c|}{\cellcolor[HTML]{F8696B}0.726}  & \multicolumn{1}{c|}{\cellcolor[HTML]{F97B6E}0.756}   & \cellcolor[HTML]{F8696B}0.726 \\ \hline
\cellcolor[HTML]{9FC5E8}SimpleLogistic               & \multicolumn{1}{c|}{\cellcolor[HTML]{FCB77A}0.852} & \multicolumn{1}{c|}{\cellcolor[HTML]{FBAC77}0.834} & \multicolumn{1}{c|}{\cellcolor[HTML]{FA8E72}0.786} & \multicolumn{1}{c|}{\cellcolor[HTML]{FBA276}0.818}   & \cellcolor[HTML]{F98D72}0.785 & \multicolumn{1}{c|}{\cellcolor[HTML]{F97D6E}0.758}  & \multicolumn{1}{c|}{\cellcolor[HTML]{F97D6E}0.758}   & \cellcolor[HTML]{F97D6E}0.758 \\ \hline
\cellcolor[HTML]{9FC5E8}SMO                          & \multicolumn{1}{c|}{\cellcolor[HTML]{FA9673}0.799} & \multicolumn{1}{c|}{\cellcolor[HTML]{F98B71}0.781} & \multicolumn{1}{c|}{\cellcolor[HTML]{F98770}0.775} & \multicolumn{1}{c|}{\cellcolor[HTML]{FBA175}0.816}   & \cellcolor[HTML]{F98770}0.774 & \multicolumn{1}{c|}{\cellcolor[HTML]{F8696B}0.727}  & \multicolumn{1}{c|}{\cellcolor[HTML]{F8766D}0.748}   & \cellcolor[HTML]{F8696B}0.726 \\ \hline
\cellcolor[HTML]{9FC5E8}Logistic                     & \multicolumn{1}{c|}{\cellcolor[HTML]{FCB77A}0.851} & \multicolumn{1}{c|}{\cellcolor[HTML]{FBAF78}0.839} & \multicolumn{1}{c|}{\cellcolor[HTML]{FA8E72}0.786} & \multicolumn{1}{c|}{\cellcolor[HTML]{FA9A74}0.805}   & \cellcolor[HTML]{FA8E72}0.786 & \multicolumn{1}{c|}{\cellcolor[HTML]{F87B6E}0.755}  & \multicolumn{1}{c|}{\cellcolor[HTML]{F87B6E}0.755}   & \cellcolor[HTML]{F87B6E}0.755 \\ \hline
\cellcolor[HTML]{B6D7A8}RandomForest                 & \multicolumn{1}{c|}{\cellcolor[HTML]{63BE7B}0.999} & \multicolumn{1}{c|}{\cellcolor[HTML]{6DC17C}0.995} & \multicolumn{1}{c|}{\cellcolor[HTML]{74C37C}0.992} & \multicolumn{1}{c|}{\cellcolor[HTML]{74C37C}0.992}   & \cellcolor[HTML]{E8E583}0.944 & \multicolumn{1}{c|}{\cellcolor[HTML]{C1DA81}0.96}   & \multicolumn{1}{c|}{\cellcolor[HTML]{FDEB84}0.935}   & \cellcolor[HTML]{BAD780}0.963 \\ \hline
\cellcolor[HTML]{B6D7A8}RandomTree                   & \multicolumn{1}{c|}{\cellcolor[HTML]{C8DC81}0.957} & \multicolumn{1}{c|}{\cellcolor[HTML]{DCE182}0.949} & \multicolumn{1}{c|}{\cellcolor[HTML]{EFE784}0.941} & \multicolumn{1}{c|}{\cellcolor[HTML]{F1E784}0.94}    & \cellcolor[HTML]{80C77D}0.987 & \multicolumn{1}{c|}{\cellcolor[HTML]{FDC77D}0.877}  & \multicolumn{1}{c|}{\cellcolor[HTML]{FDC77D}0.877}   & \cellcolor[HTML]{DCE182}0.949 \\ \hline
\cellcolor[HTML]{B6D7A8}J48                          & \multicolumn{1}{c|}{\cellcolor[HTML]{AED480}0.968} & \multicolumn{1}{c|}{\cellcolor[HTML]{94CC7E}0.979} & \multicolumn{1}{c|}{\cellcolor[HTML]{A4D17F}0.972} & \multicolumn{1}{c|}{\cellcolor[HTML]{7EC67D}0.988}   & \cellcolor[HTML]{80C77D}0.987 & \multicolumn{1}{c|}{\cellcolor[HTML]{F8E984}0.937}  & \multicolumn{1}{c|}{\cellcolor[HTML]{D2DE82}0.953}   & \cellcolor[HTML]{C6DB81}0.958 \\ \hline
\cellcolor[HTML]{B6D7A8}LMT                          & \multicolumn{1}{c|}{\cellcolor[HTML]{83C87D}0.986} & \multicolumn{1}{c|}{\cellcolor[HTML]{83C87D}0.986} & \multicolumn{1}{c|}{\cellcolor[HTML]{94CC7E}0.979} & \multicolumn{1}{c|}{\cellcolor[HTML]{77C47D}0.991}   & \cellcolor[HTML]{77C47D}0.991 & \multicolumn{1}{c|}{\cellcolor[HTML]{DCE182}0.949}  & \multicolumn{1}{c|}{\cellcolor[HTML]{D0DE82}0.954}   & \cellcolor[HTML]{C4DA81}0.959 \\ \hline
\cellcolor[HTML]{B6D7A8}REPTree                      & \multicolumn{1}{c|}{\cellcolor[HTML]{8FCB7E}0.981} & \multicolumn{1}{c|}{\cellcolor[HTML]{B0D580}0.967} & \multicolumn{1}{c|}{\cellcolor[HTML]{A4D17F}0.972} & \multicolumn{1}{c|}{\cellcolor[HTML]{79C57D}0.99}    & \cellcolor[HTML]{83C87D}0.986 & \multicolumn{1}{c|}{\cellcolor[HTML]{ECE683}0.942}  & \multicolumn{1}{c|}{\cellcolor[HTML]{C8DC81}0.957}   & \cellcolor[HTML]{C1DA81}0.96  \\ \hline
\cellcolor[HTML]{B6D7A8}HoeffdingTree                & \multicolumn{1}{c|}{\cellcolor[HTML]{FBAA77}0.831} & \multicolumn{1}{c|}{\cellcolor[HTML]{F98C71}0.783} & \multicolumn{1}{c|}{\cellcolor[HTML]{FAA075}0.815} & \multicolumn{1}{c|}{\cellcolor[HTML]{FCC47C}0.873}   & \cellcolor[HTML]{FCB379}0.846 & \multicolumn{1}{c|}{\cellcolor[HTML]{FA9172}0.79}   & \multicolumn{1}{c|}{\cellcolor[HTML]{FBA476}0.822}   & \cellcolor[HTML]{FA9E75}0.811 \\ \hline
\rowcolor[HTML]{F9806F} 
\cellcolor[HTML]{B6D7A8}DecisionStump                & \multicolumn{1}{c|}{\cellcolor[HTML]{F9806F}0.763} & \multicolumn{1}{c|}{\cellcolor[HTML]{F9806F}0.763} & \multicolumn{1}{c|}{\cellcolor[HTML]{F9806F}0.763} & \multicolumn{1}{c|}{\cellcolor[HTML]{FDD57F}0.9}     & 0.763                         & \multicolumn{1}{c|}{\cellcolor[HTML]{F9806F}0.763}  & \multicolumn{1}{c|}{\cellcolor[HTML]{FCC17C}0.868}   & 0.763                         \\ \hline
\cellcolor[HTML]{B4A7D6}JRIP                         & \multicolumn{1}{c|}{\cellcolor[HTML]{9BCE7F}0.976} & \multicolumn{1}{c|}{\cellcolor[HTML]{D2DE82}0.953} & \multicolumn{1}{c|}{\cellcolor[HTML]{D7E082}0.951} & \multicolumn{1}{c|}{\cellcolor[HTML]{85C87D}0.985}   & \cellcolor[HTML]{7CC57D}0.989 & \multicolumn{1}{c|}{\cellcolor[HTML]{FDD680}0.902}  & \multicolumn{1}{c|}{\cellcolor[HTML]{D4DF82}0.952}   & \cellcolor[HTML]{D7E082}0.951 \\ \hline
\cellcolor[HTML]{B4A7D6}DecisionTable                & \multicolumn{1}{c|}{\cellcolor[HTML]{9BCE7F}0.976} & \multicolumn{1}{c|}{\cellcolor[HTML]{C8DC81}0.957} & \multicolumn{1}{c|}{\cellcolor[HTML]{C6DB81}0.958} & \multicolumn{1}{c|}{\cellcolor[HTML]{8CCA7E}0.982}   & \cellcolor[HTML]{A7D27F}0.971 & \multicolumn{1}{c|}{\cellcolor[HTML]{FFEB84}0.934}  & \multicolumn{1}{c|}{\cellcolor[HTML]{DCE182}0.949}   & \cellcolor[HTML]{EAE583}0.943 \\ \hline
\cellcolor[HTML]{B4A7D6}Part                         & \multicolumn{1}{c|}{\cellcolor[HTML]{94CC7E}0.979} & \multicolumn{1}{c|}{\cellcolor[HTML]{98CE7F}0.977} & \multicolumn{1}{c|}{\cellcolor[HTML]{A7D27F}0.971} & \multicolumn{1}{c|}{\cellcolor[HTML]{77C47D}0.991}   & \cellcolor[HTML]{83C87D}0.986 & \multicolumn{1}{c|}{\cellcolor[HTML]{FFEB84}0.934}  & \multicolumn{1}{c|}{\cellcolor[HTML]{E3E383}0.946}   & \cellcolor[HTML]{D0DE82}0.954 \\ \hline
\cellcolor[HTML]{B4A7D6}OneR                         & \multicolumn{1}{c|}{\cellcolor[HTML]{FBAE78}0.838} & \multicolumn{1}{c|}{\cellcolor[HTML]{FBAE78}0.838} & \multicolumn{1}{c|}{\cellcolor[HTML]{FBAE78}0.838} & \multicolumn{1}{c|}{\cellcolor[HTML]{FEEA83}0.933}   & \cellcolor[HTML]{FEDD81}0.913 & \multicolumn{1}{c|}{\cellcolor[HTML]{FBAE78}0.838}  & \multicolumn{1}{c|}{\cellcolor[HTML]{FEDA80}0.907}   & \cellcolor[HTML]{FEDD81}0.913 \\ \hline
\cellcolor[HTML]{FFE599}BayesNet                     & \multicolumn{1}{c|}{\cellcolor[HTML]{9BCE7F}0.976} & \multicolumn{1}{c|}{\cellcolor[HTML]{E0E383}0.947} & \multicolumn{1}{c|}{\cellcolor[HTML]{F1E784}0.94}  & \multicolumn{1}{c|}{\cellcolor[HTML]{EFE784}0.941}   & \cellcolor[HTML]{E8E583}0.944 & \multicolumn{1}{c|}{\cellcolor[HTML]{FEE182}0.919}  & \multicolumn{1}{c|}{\cellcolor[HTML]{FEE082}0.918}   & \cellcolor[HTML]{FEE683}0.927 \\ \hline
\cellcolor[HTML]{F9CB9C}KNN                          & \multicolumn{1}{c|}{\cellcolor[HTML]{A0D07F}0.974} & \multicolumn{1}{c|}{\cellcolor[HTML]{C8DC81}0.957} & \multicolumn{1}{c|}{\cellcolor[HTML]{E0E383}0.947} & \multicolumn{1}{c|}{\cellcolor[HTML]{E0E383}0.947}   & \cellcolor[HTML]{A7D27F}0.971 & \multicolumn{1}{c|}{\cellcolor[HTML]{FDD07E}0.892}  & \multicolumn{1}{c|}{\cellcolor[HTML]{FDD07E}0.892}   & \cellcolor[HTML]{EAE583}0.943 \\ \hline
\end{tabular}}
\end{table}

To quantify the overall performance of the HMD classifier, the combination of both accuracy and area under the receiver operating characteristics curve is taken into account to capture both the general correctness and discriminative ability of a hardware malware detector classifier. By multiplying these two, we will reward the classifiers with both good accuracy and discriminative ability while penalizing those that perform poorly in either metric. We define this metric of performance as:

\begin{equation}
\text{Performance} = \text{Accuracy} \times \text{AUC(ROC)}
\end{equation}

This ensures that a classifier only achieves a high performance score if it is both accurate in its predictions across benign and malware classes and effective at distinguishing between positive and negative instances across various thresholds. This combination makes the performance metric more robust than using either accuracy or AUC(ROC) alone, especially in this scenario where class imbalances might skew the perceived effectiveness of the classifier when judged by accuracy alone. Table VI displays the calculated values for the specified performance measure.
% \usepackage{multirow}
\begin{table}[]
\centering
\caption{A Comparison of classifiers based on the Performance metric}
\resizebox{\columnwidth}{!}{%
\begin{tabular}{|c|ccccc|ccc|}
\hline
\rowcolor[HTML]{D5A6BD} 
\cellcolor[HTML]{D5A6BD}                             & \multicolumn{1}{c|}{\cellcolor[HTML]{D5A6BD}HPC16} & \multicolumn{1}{c|}{\cellcolor[HTML]{D5A6BD}HPC8}  & \multicolumn{3}{c|}{\cellcolor[HTML]{D5A6BD}HPC4}                                                                                         & \multicolumn{3}{c|}{\cellcolor[HTML]{D5A6BD}HPC2}                                                                                          \\ \cline{2-9} 
\rowcolor[HTML]{D5A6BD} 
\multirow{-2}{*}{\cellcolor[HTML]{D5A6BD}Classifier} & \multicolumn{3}{c|}{\cellcolor[HTML]{D5A6BD}Single}                                                                                                          & \multicolumn{1}{c|}{\cellcolor[HTML]{D5A6BD}Boosted} & Bagging                       & \multicolumn{1}{c|}{\cellcolor[HTML]{D5A6BD}Single} & \multicolumn{1}{c|}{\cellcolor[HTML]{D5A6BD}Boosted} & Bagging                       \\ \hline
\cellcolor[HTML]{9FC5E8}MLP                          & \multicolumn{1}{c|}{\cellcolor[HTML]{FEE582}79.05} & \multicolumn{1}{c|}{\cellcolor[HTML]{FDCA7D}73.37} & \multicolumn{1}{c|}{\cellcolor[HTML]{FBAD78}67.20} & \multicolumn{1}{c|}{\cellcolor[HTML]{FCBF7B}70.92}   & \cellcolor[HTML]{FCC27C}71.50 & \multicolumn{1}{c|}{\cellcolor[HTML]{FA8E72}60.50}  & \multicolumn{1}{c|}{\cellcolor[HTML]{FA8E72}60.50}   & \cellcolor[HTML]{FA9473}61.87 \\ \hline
\cellcolor[HTML]{9FC5E8}SGD                          & \multicolumn{1}{c|}{\cellcolor[HTML]{FA9F75}64.12} & \multicolumn{1}{c|}{\cellcolor[HTML]{FA9473}61.79} & \multicolumn{1}{c|}{\cellcolor[HTML]{F98C71}60.05} & \multicolumn{1}{c|}{\cellcolor[HTML]{FA9D75}63.76}   & \cellcolor[HTML]{F98D71}60.32 & \multicolumn{1}{c|}{\cellcolor[HTML]{F8696B}52.59}  & \multicolumn{1}{c|}{\cellcolor[HTML]{F8706C}54.08}   & \cellcolor[HTML]{F8696B}52.68 \\ \hline
\cellcolor[HTML]{9FC5E8}SimpleLogistic               & \multicolumn{1}{c|}{\cellcolor[HTML]{FBB078}67.70} & \multicolumn{1}{c|}{\cellcolor[HTML]{FBA275}64.73} & \multicolumn{1}{c|}{\cellcolor[HTML]{F98B71}59.87} & \multicolumn{1}{c|}{\cellcolor[HTML]{FA9673}62.31}   & \cellcolor[HTML]{F98A71}59.72 & \multicolumn{1}{c|}{\cellcolor[HTML]{F8706C}54.22}  & \multicolumn{1}{c|}{\cellcolor[HTML]{F8706C}54.22}   & \cellcolor[HTML]{F8706C}54.15 \\ \hline
\cellcolor[HTML]{9FC5E8}SMO                          & \multicolumn{1}{c|}{\cellcolor[HTML]{FA9D75}63.67} & \multicolumn{1}{c|}{\cellcolor[HTML]{FA8F72}60.86} & \multicolumn{1}{c|}{\cellcolor[HTML]{F98B71}59.91} & \multicolumn{1}{c|}{\cellcolor[HTML]{FA9B74}63.37}   & \cellcolor[HTML]{F98A71}59.79 & \multicolumn{1}{c|}{\cellcolor[HTML]{F8696B}52.79}  & \multicolumn{1}{c|}{\cellcolor[HTML]{F8716C}54.31}   & \cellcolor[HTML]{F8696B}52.72 \\ \hline
\cellcolor[HTML]{9FC5E8}Logistic                     & \multicolumn{1}{c|}{\cellcolor[HTML]{FBAF78}67.58} & \multicolumn{1}{c|}{\cellcolor[HTML]{FBA576}65.42} & \multicolumn{1}{c|}{\cellcolor[HTML]{F98B71}59.87} & \multicolumn{1}{c|}{\cellcolor[HTML]{FA9272}61.32}   & \cellcolor[HTML]{F98B71}59.91 & \multicolumn{1}{c|}{\cellcolor[HTML]{F86C6B}53.43}  & \multicolumn{1}{c|}{\cellcolor[HTML]{F86C6B}53.43}   & \cellcolor[HTML]{F86C6B}53.43 \\ \hline
\cellcolor[HTML]{B6D7A8}RandomForest                 & \multicolumn{1}{c|}{\cellcolor[HTML]{63BE7B}98.23} & \multicolumn{1}{c|}{\cellcolor[HTML]{71C37C}96.63} & \multicolumn{1}{c|}{\cellcolor[HTML]{81C77D}94.87} & \multicolumn{1}{c|}{\cellcolor[HTML]{81C77D}94.82}   & \cellcolor[HTML]{A9D27F}90.23 & \multicolumn{1}{c|}{\cellcolor[HTML]{D1DE82}85.62}  & \multicolumn{1}{c|}{\cellcolor[HTML]{E5E483}83.27}   & \cellcolor[HTML]{D1DE82}85.63 \\ \hline
\cellcolor[HTML]{B6D7A8}RandomTree                   & \multicolumn{1}{c|}{\cellcolor[HTML]{9DCF7F}91.56} & \multicolumn{1}{c|}{\cellcolor[HTML]{ABD380}90.03} & \multicolumn{1}{c|}{\cellcolor[HTML]{B7D780}88.59} & \multicolumn{1}{c|}{\cellcolor[HTML]{B9D780}88.33}   & \cellcolor[HTML]{82C77D}94.70 & \multicolumn{1}{c|}{\cellcolor[HTML]{FEDB81}76.92}  & \multicolumn{1}{c|}{\cellcolor[HTML]{FEDB81}76.92}   & \cellcolor[HTML]{DFE283}83.96 \\ \hline
\cellcolor[HTML]{B6D7A8}J48                          & \multicolumn{1}{c|}{\cellcolor[HTML]{8DCA7E}93.49} & \multicolumn{1}{c|}{\cellcolor[HTML]{87C97E}94.11} & \multicolumn{1}{c|}{\cellcolor[HTML]{A0D07F}91.25} & \multicolumn{1}{c|}{\cellcolor[HTML]{84C87D}94.53}   & \cellcolor[HTML]{8ACA7E}93.77 & \multicolumn{1}{c|}{\cellcolor[HTML]{EDE683}82.39}  & \multicolumn{1}{c|}{\cellcolor[HTML]{E3E383}83.45}   & \cellcolor[HTML]{D2DE82}85.44 \\ \hline
\cellcolor[HTML]{B6D7A8}LMT                          & \multicolumn{1}{c|}{\cellcolor[HTML]{7CC67D}95.40} & \multicolumn{1}{c|}{\cellcolor[HTML]{84C87D}94.43} & \multicolumn{1}{c|}{\cellcolor[HTML]{95CD7E}92.48} & \multicolumn{1}{c|}{\cellcolor[HTML]{83C77D}94.64}   & \cellcolor[HTML]{86C87D}94.23 & \multicolumn{1}{c|}{\cellcolor[HTML]{D9E082}84.64}  & \multicolumn{1}{c|}{\cellcolor[HTML]{DFE283}83.97}   & \cellcolor[HTML]{D4DF82}85.23 \\ \hline
\cellcolor[HTML]{B6D7A8}REPTree                      & \multicolumn{1}{c|}{\cellcolor[HTML]{87C97E}94.17} & \multicolumn{1}{c|}{\cellcolor[HTML]{9DCF7F}91.60} & \multicolumn{1}{c|}{\cellcolor[HTML]{A0D07F}91.20} & \multicolumn{1}{c|}{\cellcolor[HTML]{85C87D}94.36}   & \cellcolor[HTML]{90CB7E}93.09 & \multicolumn{1}{c|}{\cellcolor[HTML]{E1E383}83.68}  & \multicolumn{1}{c|}{\cellcolor[HTML]{D7E082}84.88}   & \cellcolor[HTML]{D3DF82}85.41 \\ \hline
\cellcolor[HTML]{B6D7A8}HoeffdingTree                & \multicolumn{1}{c|}{\cellcolor[HTML]{FBA776}65.81} & \multicolumn{1}{c|}{\cellcolor[HTML]{F98D71}60.28} & \multicolumn{1}{c|}{\cellcolor[HTML]{FA9874}62.74} & \multicolumn{1}{c|}{\cellcolor[HTML]{FBAB77}66.69}   & \cellcolor[HTML]{FBA777}65.93 & \multicolumn{1}{c|}{\cellcolor[HTML]{F98C71}60.07}  & \multicolumn{1}{c|}{\cellcolor[HTML]{FA9773}62.39}   & \cellcolor[HTML]{FA9373}61.67 \\ \hline
\rowcolor[HTML]{F9826F} 
\cellcolor[HTML]{B6D7A8}DecisionStump                & \multicolumn{1}{c|}{\cellcolor[HTML]{F9826F}58.02} & \multicolumn{1}{c|}{\cellcolor[HTML]{F9826F}58.02} & \multicolumn{1}{c|}{\cellcolor[HTML]{F9826F}58.02} & \multicolumn{1}{c|}{\cellcolor[HTML]{FDD07E}74.64}   & 58.02                         & \multicolumn{1}{c|}{\cellcolor[HTML]{F9826F}58.02}  & \multicolumn{1}{c|}{\cellcolor[HTML]{FBA877}66.00}   & 58.02                         \\ \hline
\cellcolor[HTML]{B4A7D6}JRIP                         & \multicolumn{1}{c|}{\cellcolor[HTML]{89C97E}93.91} & \multicolumn{1}{c|}{\cellcolor[HTML]{AED480}89.68} & \multicolumn{1}{c|}{\cellcolor[HTML]{B9D780}88.33} & \multicolumn{1}{c|}{\cellcolor[HTML]{96CD7E}92.38}   & \cellcolor[HTML]{8DCA7E}93.46 & \multicolumn{1}{c|}{\cellcolor[HTML]{FEE382}78.70}  & \multicolumn{1}{c|}{\cellcolor[HTML]{E1E383}83.71}   & \cellcolor[HTML]{DFE283}84.00 \\ \hline
\cellcolor[HTML]{B4A7D6}DecisionTable                & \multicolumn{1}{c|}{\cellcolor[HTML]{8ECB7E}93.34} & \multicolumn{1}{c|}{\cellcolor[HTML]{ACD380}89.84} & \multicolumn{1}{c|}{\cellcolor[HTML]{BAD780}88.25} & \multicolumn{1}{c|}{\cellcolor[HTML]{97CD7E}92.27}   & \cellcolor[HTML]{B2D580}89.14 & \multicolumn{1}{c|}{\cellcolor[HTML]{EFE784}82.12}  & \multicolumn{1}{c|}{\cellcolor[HTML]{E2E383}83.66}   & \cellcolor[HTML]{E9E583}82.79 \\ \hline
\cellcolor[HTML]{B4A7D6}Part                         & \multicolumn{1}{c|}{\cellcolor[HTML]{85C87D}94.33} & \multicolumn{1}{c|}{\cellcolor[HTML]{A0D07F}91.27} & \multicolumn{1}{c|}{\cellcolor[HTML]{ADD480}89.80} & \multicolumn{1}{c|}{\cellcolor[HTML]{83C77D}94.64}   & \cellcolor[HTML]{95CD7E}92.47 & \multicolumn{1}{c|}{\cellcolor[HTML]{FAEA84}80.82}  & \multicolumn{1}{c|}{\cellcolor[HTML]{E6E483}83.14}   & \cellcolor[HTML]{DBE182}84.44 \\ \hline
\cellcolor[HTML]{B4A7D6}OneR                         & \multicolumn{1}{c|}{\cellcolor[HTML]{FCBC7A}70.25} & \multicolumn{1}{c|}{\cellcolor[HTML]{FCBC7A}70.25} & \multicolumn{1}{c|}{\cellcolor[HTML]{FCBC7A}70.25} & \multicolumn{1}{c|}{\cellcolor[HTML]{FEE883}79.60}   & \cellcolor[HTML]{FEDD81}77.32 & \multicolumn{1}{c|}{\cellcolor[HTML]{FCBC7A}70.25}  & \multicolumn{1}{c|}{\cellcolor[HTML]{FDD47F}75.46}   & \cellcolor[HTML]{FEDD81}77.32 \\ \hline
\cellcolor[HTML]{FFE599}BayesNet                     & \multicolumn{1}{c|}{\cellcolor[HTML]{89C97E}93.91} & \multicolumn{1}{c|}{\cellcolor[HTML]{D7E082}84.93} & \multicolumn{1}{c|}{\cellcolor[HTML]{EEE784}82.19} & \multicolumn{1}{c|}{\cellcolor[HTML]{F1E784}81.85}   & \cellcolor[HTML]{EEE784}82.20 & \multicolumn{1}{c|}{\cellcolor[HTML]{FEE082}78.07}  & \multicolumn{1}{c|}{\cellcolor[HTML]{FEDE81}77.49}   & \cellcolor[HTML]{FEE382}78.63 \\ \hline
\cellcolor[HTML]{F9CB9C}KNN                          & \multicolumn{1}{c|}{\cellcolor[HTML]{81C77D}94.86} & \multicolumn{1}{c|}{\cellcolor[HTML]{9DCF7F}91.60} & \multicolumn{1}{c|}{\cellcolor[HTML]{AED480}89.67} & \multicolumn{1}{c|}{\cellcolor[HTML]{AED480}89.67}   & \cellcolor[HTML]{9ACE7F}91.90 & \multicolumn{1}{c|}{\cellcolor[HTML]{FEE783}79.52}  & \multicolumn{1}{c|}{\cellcolor[HTML]{FEE783}79.52}   & \cellcolor[HTML]{E1E383}83.68 \\ \hline
\end{tabular}}
\end{table}

\subsection{Experimental Results }
The multi-layer perceptron has the best performance in function-based classifiers; its performance improves with both the boosting and bagging methods for 4-HPC and with the bagging method for the 2-HPC configuration. The SGD classifier improves while boosted and keeps the same performance for the bagging technique. The Simple Logistic improves with the boosted technique for 4-HPC and with both boosted and bagging for the 2-HPC configuration.

For the tree-based classifiers, the Random Forest has the best performance; it improves for both the boosted and bagging techniques, while for the boosted method in the case of 2 HPCs it loses performance. The Random Tree classifier keeps the same performance for boosting and improves by 6\% for bagging in the 4-HPC configuration, but no improvement is observed in the case of 2-HPC ensemble techniques. The J48 is improved by 4\% and 3\% for the boosted and bagging strategies, respectively. Its performance increases by 1\% and 2\% for the case of the 2-HPC configuration as well.

The JRIP classifier improves by 5\% and 6\% for the boosted and bagging techniques in both the 4-HPC and 2-HPC mode configurations, respectively. The Decision Table classifier improves its performance by 4\% in the boosted technique and 1\% in the bagging technique; however, in the case of 2-HPC, a performance improvement is not found. The Part classifier increases by 5\% and 3\% for the boosted and bagging methods in 4-HPC mode, respectively. The performance also improves by 4\% and 3\% for the 2-HPC mode configuration.

Since the other candidates in these two families didn’t show a minimum threshold in the performance even with the 16 HPC configuration, we could only shortlist the Bayes Net and the KNN classifier from those families. The Bayes Net loses performance with the boosted technique for both 4 and 2 HPC modes, and keeps the same performance for the bagging methods in both configurations. The only improvement in the KNN is observed in the bagging method with the 2-HPC configuration, which is near 4\%.

Considering both the performance and area under the ROC curve, we ranked all the classifiers with performance metric as mentioned above. The Random Forest, LMT and J48 are performing better than all other 15 classifiers as shown in Table VI. Also, it is important to note that the performance drops severely in the transition from the 4-HPC configuration to the 2-HPC configuration. Hence, from a malware detector designer's perspective, we recommend using at least 4 HPCs and using or skipping the ensemble methods based on the individual classifier results provided above.

%%FFF Fig5
%\begin{figure}[htbp]
%\centerline{\includegraphics[width=\columnwidth]%{Figures/Figure5.png}}
%\caption{Performance of function-based classifiers}
%\label{fig}
%\end{figure}

%%FFF Fig6
%\begin{figure}[htbp]
%\centerline{\includegraphics[width=\columnwidth]%{Figures/Figure6.png}}
%\caption{Performance of tree-based classifiers}
%\label{fig}
%\end{figure}

%%FFF Fig7
%\begin{figure}[htbp]
%\centerline{\includegraphics[width=\columnwidth]%{Figures/Figure7.png}}
%\caption{Performance of rule-based classifiers}
%\label{fig}
%\end{figure}

%FFF Fig8
%\begin{figure}[htbp]
%\centerline{\includegraphics[width=\columnwidth]{Figures/Figure8.png}}
%\caption{Performance of Bayes and Lazy classifiers}
%\label{fig}
%\end{figure}

\section{Conclusions}

In this research, we examined multiple machine learning models for classifying hardware performance counters (HPC) to detect malware. Our results demonstrate that these models effectively distinguish between malicious and benign applications, providing a viable defense against increasingly complex malware. We then applied a ranking algorithm to select and weight HPC features from a reduced set of features. Features such as L1-dcache-loads and L1-dcache-stores were especially effective in distinguishing malware from benign applications. We evaluated malware detectors with 144 experiments using different machine learning models, including ensemble methods such as boosting and bagging. Our results demonstrate that linear classifiers are inappropriate for malware detection, as hardware footprints in HPCs are nonlinear. We found that ensemble configurations of the Random Tree, J48, REP Tree, Decision Stump, Part, and OneR classifiers outperformed their simpler counterparts with 16 HPCs, using four HPCs. This could mean a 75\% reduction in input counts while keeping the same performance, which means a very adaptable hardware malware detector. Additionally, ensemble models composed of SGD, SMO, LMT, Hoeffding Tree, JRIP, Decision Table, and KNN classifiers, optimized with only four attributes, performed better than their simpler versions with eight HPCs, allowing for a 50\% reduction in inputs without sacrificing performance in a general-purpose safe execution environment. However, reducing feature attributes in MLP, Simple Logistic, Logistic, Random Forest, BayesNet, and using ensemble methods did not improve the performance scores.

\begingroup
\renewcommand{\footnotesize}{\fontsize{8pt}{8.7pt}\selectfont}

\endgroup
\end{document}